# Impact of impurities on leakage current induced by High-Energy Density Pulsed Laser Annealing in Si diodes

Richard Monflier, Richard Daubriac*, Mahmoud Haned, Toshiyuki Tabata, François Olivier, Eric Imbernon, Markus Italia, Antonino La Magna, Fulvio Mazzamuto, Simona Boninelli, Fuccio Cristiano, Elena Bedel Pereira

*Abstract*—For semiconductor device fabrication, Pulsed Laser Annealing (PLA) offers significant advantages over conventional thermal processes. Notably, it can provide ultrafast (~ns) and high temperature profiles (>1000°C). When the maximum temperature exceeds the melting point, a solid-liquid phase transition is observed, immediately followed by rapid recrystallization. This unique annealing mechanism gives raises questions about dopant diffusion and residual defects, in not only in the recrystallized region, but also just below it. As power devices require micrometer-sized junctions, high laser energy densities are needed, which were proved to promote the incorporation of complex impurities from the surface and the creation of defects at the liquid/solid interface. This paper reports on the impact of laser annealing at high energy densities (up to 8.0 J/cm²) on the leakage current, using Schottky and PN diodes, and DLTS measurements. Various laser annealing conditions were used: energy densities between 1.7 and 8.0 J/cm² with 1 to 10 pulses. Our results suggest that the liquid and solid solubility of vacancies in silicon are fixed by the maximum temperature reached, so to the energy density. Increasing the number of laser pulses allows, not only to reach this maximum vacancy concentration but also to promote their diffusion towards the surface. Concomitantly, the in-diffusion of complex impurities inside the melted region allows the coupling between both defect types to create trap centers, responsible for the degradation of the leakage current.

*Index Terms*—Pulsed laser annealing, leakage current, complex impurities, vacancies.

Date of submission: 09/01/2025
This work was financially supported by the Nano2017 French program and was partly supported by LAAS-CNRS micro and nanotechnologies platform, member of the French RENATECH network. We dedicate this article in loving memory of our friend and colleague Fuccio Cristiano, who passed away in January 2024, for his extensive contribution to this work and to other related projects.

Richard Monflier, Richard Daubriac*, Mahmoud Haned, François Olivier, Eric Imbernon, Fuccio Cristiano and Elena Bedel Pereira are with LAAS-CNRS, CNRS, Université de Toulouse, UPS Toulouse, 31400, France (e-mail: richard.daubriac@laas.fr).
Toshiyuki Tabata and Fulvio Mazzamuto are with Laser Systems & Solutions of Europe (LASSE), Gennevilliers, France.
Markus Italia, Antonino La Magna and Simona Boninelli are with CNR-IMM, Catania (Italy).

## I. INTRODUCTION

OVER the past decades, Pulsed Laser Annealing (PLA) has been identified as one of the most promising candidates for device fabrication. Among many advantages, it enables fast (~ns) and high-temperatures (>1000°C) processing. By using low energy density conditions, this type of budget can be applied at the near surface (~100 nm). This significant and confined thermal budget promotes the out-of-equilibrium phenomenon of rapid recrystallization occurring in locally melted regions and enables the solid solubility of dopants to be exceeded [1,2]. These properties are beneficial for the fabrication of ultra-doped shallow junctions and the design of future transistors, especially in the 3D monolithic architecture [3-6]. However, it has also been demonstrated that PLA can involve significant crystalline damage [7], leading to electrical degradation such as a reduction in carrier lifetime [8], an increase of leakage current [9,10]. For power devices, micrometer-thick junctions are required, which necessitates the use of high energy densities (> 4.0 J/cm²). At the same time as increasing the melt depth, the use of such laser annealing conditions leads to an increase in the duration of the liquid phase. Consequently, in addition to the issues observed at low energy density laser conditions, other undesirable phenomena may occur, thus leading degradation of the electrical properties of the device.

Previous studies on defect formation induced by laser annealing have identified two main types of defects: impurity complexes (carbon and oxygen) [11] and point defects (vacancies) [12,13]. Impurity complexes are incorporated from the ambient environment into the liquid phase and in-diffuse from the surface down to the liquid/solid (l/s) interface. On the opposite, vacancies arise at the l/s interface and their diffusion occurs on both sides of this interface. In the liquid region, they diffuse towards the surface while, in the solid region, they accumulate near the interface. The evolution of their depth profiles was studied by varying the laser annealing conditions in energy density (ED) and number of pulses ($N_P$) (see Fig. 1). For complex impurities, it was shown that their depth-concentration profiles increase in concentration values at the surface with increasing $N_P$ and extend closer and closer to the l/s interface. When the ED is increased, the same behavior is observed with the l/s interface shifting deeper into the material as the only difference. For vacancies, increasing the ED pushes the l/s deeper in the material without involving significant changes whereas increasing $N_P$ generates more vacancies, that accumulate on the solid side of the l/s interface.

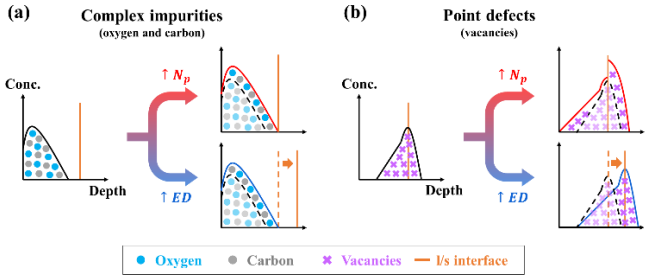

Fig. 1. Graphical representation of (a) the complex impurities (carbon and oxygen) and (b) point defects (vacancies) depth profiles after 1 pulse and their evolution when applying increasing laser energy density (ED) or number of pulses ($N_P$) [11-13].

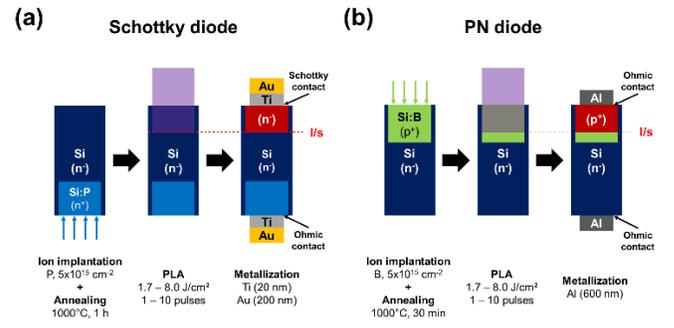

Fig. 2. Schematics of the (a) Schottky and (b) PN diodes structures with their respective fabrication process.

In this paper, we have investigated the impact of such impurities and defects, generated by high ED and multi-pulse laser annealing, on the leakage current. This study was conducted using three main characterization methods: spreading resistance profiles (SRP), current-voltage characteristics (I-V) measurements with dedicated Schottky and PN diode structures, and deep level transient spectroscopy (DLTS).

## II. EXPERIMENT

### A. Fabrication process

In this work, two types of structures have been fabricated: Schottky and PN diodes [14-18]. Schottky diodes allow the characterization of the impact of complex impurities and point defects within the melted region (from the surface), whereas the PN diodes are focused in the region around the liquid/solid interface (l/s). Schottky diodes of different sizes (25, 35, 50, 75, 200 and 225 μm) were designed. The fabrication process (Fig. 2(a)) was achieved on two separate 100 mm n-type Si(100) wafers (Czochralski, 5 Ω.cm). The first step consisted in phosphorus implantation ($5.0 \times 10^{15}$ cm$^{-2}$, 50 keV) on the back side of the wafers, followed by a conventional activation annealing (1000 °C, 1 h, $N_2$). In the second step, PLA was applied on the front side of the wafers under different conditions. Finally, a Ti(20 nm)/Au(200 nm) metal deposition was performed on the front and back sides to create Schottky and ohmic contacts, respectively (immediately after RCA cleaning: SC1, SC2 and HF). Prior to PLA, one wafer was chemically etched to remove the native oxide using BOE (Buffered Oxide Etch) and thermally oxidized to form a 4.0 nm-thick silicon dioxide. No treatment was applied to the other wafer. For the sake of clarity, these wafers are referred to as "thermal" and "standard", respectively. After the PLA process, both wafers were cleaned with BOE to remove native and thermal oxides. The PN diodes fabrication process (Fig. 2(b)) was performed on a 100 mm n-type Si(100) wafer (Czochralski, 1.3 Ω.cm). The first step consisted in boron implantation ($5.0 \times 10^{15}$ cm$^{-2}$, 50 keV) on the front side of the wafer, followed by an activation annealing (1000 °C, 30 min, $N_2$). This annealing allows a boron diffusion of 1.4 μm in the silicon.

During the second step, PLA was applied on the front of the wafers at various conditions. Finally, aluminum metal deposition (600 nm) was performed on both sides of the wafer to create ohmic contacts. For both diode structures, the PLA process was performed using a LT3100 system from SCREEN-LASSE, equipped with a XeCl excimer laser (308 nm, 160 ns). Irradiated areas of 10 x 10 mm² received ED ranging from 1.7 to 8.0 J/cm² with 1 to 10 pulses, at room temperature and under $N_2$ flux (total of 45 conditions).

### B. Characterization methods

Three main characterization methods were used in this study: Spreading Resistance Profiling (SRP) [19], I-V measurements and Deep Level Transient Spectroscopy (DLTS) [20,21]. SRP measurements were performed using a Solid-State Measurements SSM 150 system to probe the depth-concentration profiles. A bevel was prepared on samples annealed between 1.7 and 8.0 J/cm² with 10 pulses to obtain a profile resolution of 7 nm. I-V measurements were performed using a Karl Suss PA200 probe station and an Agilent 4142B parametric tester consisting of six source/measure units (SMU) to characterize the leakage current. During the measurement, the diode structure is positioned on a nickel support and connected to the ground by a dedicated SMU. The compliance voltage and current are set to 1000 V and 10 mA, respectively. The DLTS technique was used to identify the signature of the impurities and their spectra were acquired with a Bio-Rad DL8000 measurement system, equipped with a liquid $N_2$ cryostat (0.9 mbar). The temperature range of interest is between 85 and 550 K. The samples are polarized via a capacitor by an internal generator, which is tunable via a voltage source (range: -100 to +100 V, minimum step of 1 mV). A voltage pulse is described by its duration $T_p$ and its amplitude $V_p$. The low level of the pulse corresponds to voltage $V_r$. The emission of carriers trapped by defects is detected at the end of the voltage pulse and for a time duration $T_w$. The Boonton 72B capacitor operates with an alternating signal (100 mV, 1.0 MHz). The processing is done according to the Weiss method [22].

## III. RESULTS AND DISCUSSION

### A. Impact of laser annealing on electrically active complex impurities

Among the incorporated impurities, it is imperative to evaluate the electrically active portion in order to anticipate its impact on a potentially intentional doping process. Such analysis has been performed by measuring the depth profiles of the active carrier concentration using the SRP technique. The active profiles obtained for the "standard" samples annealed with 10 pulses are shown in Fig. 3. As observed, all measured profiles are n-type. Due to their electronic configuration, carbon atoms are not expected to contribute to the electrical transport, unlike oxygen atoms which act as donors when incorporated in substitution sites [23]. Therefore, we can assume that n-type profiles of SRP are due to oxygen doping. For this reason, we chose to compare these active profiles with the oxygen chemical SIMS profiles corresponding to the same or equivalent laser annealing conditions.

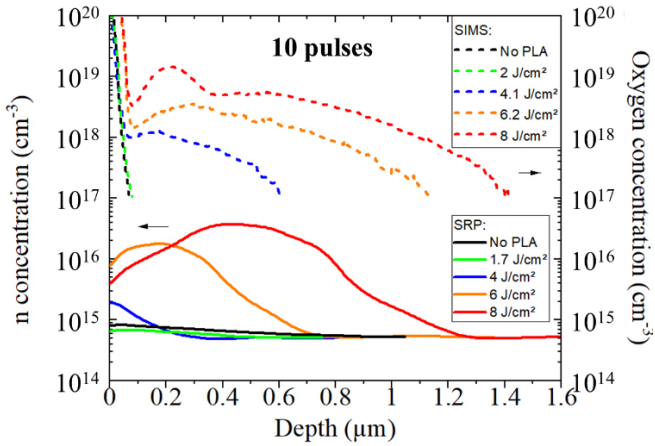

Fig. 3. SRP active n-type carrier and SIMS chemical (oxygen) concentration depth profiles obtained on "standard" samples laser-annealed at ED between 1.7 and 8.0 J/cm² with 10 pulses, along with the reference sample (no PLA).

TABLE I
ACTIVATION RATE OF THE "STANDARD" SAMPLES LASER-ANNEALED AT ED BETWEEN 1.7 AND 8.0 J/CM² WITH 10 PULSES, ALONG WITH THE REFERENCE SAMPLE (NO-PLA)

| ED (J/cm²) | 0 | 1.7 | 4.0 | 6.0 | 8.0 |
|---|---|---|---|---|---|
| Ω (%) | 0 | ~ 0 | 0.04 | 0.33 | 0.33 |

Compared to the reference sample (no PLA), we observe that the carrier concentration profiles are not significantly modified up to 4.0 J/cm². At this condition, we observe a concentration peak up to a value of $2 \times 10^{15}$ cm⁻³, which decreases towards a plateau of $7.0 \times 10^{14}$ cm⁻³ at 200 nm, similar to the no-PLA sample. For higher ED at 6.0 and 8.0 J/cm², the measured concentration profiles are Gaussian-shaped with a maximum of $1.5 \times 10^{16}$ cm⁻³ at 160 nm and $3 \times 10^{17}$ cm⁻³ at 430 nm, respectively. Also, for these two conditions, the no-LTA concentration plateau is recovered at depths of 770 and 1300 nm, respectively. These results emphasize that the application of laser annealing at an ED higher than 4.0 J/cm² modifies the initial n-type carrier concentration of the Si substrate.

For the "thermal" samples, Secondary-Ion Mass Spectrometry (SIMS) and Photo-Luminescence Spectroscopy (PLS) measurements showed oxygen in-diffusion in the molten Si [11]. This phenomenon is amplified when increasing the ED. Contrary to carbon, the introduction of oxygen impurities into substitution sites acts as a donor doping, increasing the initial n-type carrier concentration [23]. By comparing the chemical and electrical doses of oxygen extracted from the SIMS and SRP profiles, respectively, the activation rate Ω of oxygen impurities can be deduced (Table 1). This rate is found to be close to 0 for unannealed and annealed samples at 1.7 and 4.0 J/cm². At ED of 6.0 J/cm² and above, Ω increases up to ~ 0.3%, which is high enough to impact the electrical properties of devices, especially when p-doped.

### B. Impact of laser annealing on leakage current

Most of the impurities and defects we found are electrically neutral, but may have an impact on the leakage current. We used I-V measurements on fabricated Schottky and PN diodes to evaluate the impact of the PLA process on the leakage current.

*1) Schottky diodes measurements*

Fig. 4(a) summarizes the current-voltage characteristics of the "standard" samples annealed with 10 pulses (ED between 1.7 and 8.0 J/cm²) and the reference sample (No-PLA) obtained from 225 μm radius Schottky diodes, at room temperature. It should be noted that similar results were obtained for all diode sizes. We observe that, for all laser conditions, the I-V characteristics in forward bias remain unchanged, with a voltage threshold of ~ 0.1 V ± 0.5%. In reverse bias, we observe an increase in current density for ED higher than 1.7 J/cm². At 4.0 J/cm², the trend of the I-V curve is similar to the reference and 1.7 J/cm² ones but with an additional current offset of around 0.1 A/cm². For the 6.0 and 8.0 J/cm² conditions, a continuous increase is observed as the reverse voltage is increased up to approximately 1.0 A/cm². The progressive increase in current density with increasing ED can be explained by a concomitant increase in leakage current. This phenomenon is particularly pronounced at ED higher than 4.0 J/cm² and could be explained by the incorporation of additional impurities such as point defects (vacancies) within the melt depth.

From -4 V to -10 V, we notice that the current density measured at 6.0 J/cm² is significantly higher than at 8.0 J/cm². This behavior could be related to the space charge region width W (probed depth). For a Schottky diode, it can be calculated from (1) [15].

$$W = \sqrt{\frac{2\varepsilon_S(V_{bi}-V_A)}{qN_D}} \qquad (1)$$

with the dielectric constant $\varepsilon_S$, the built-in voltage $V_{bi}$, the applied voltage $V_A$, the elementary charge q and the donor concentration $N_D$. Since the depth concentration profile $N_D$ is defined by the laser annealing process, the application of a voltage V allows to tune the depth that is electrically probed by modifying the space charge region width W. In order to properly examine the electrical properties of our films, W must be compared with the melt depth $D_m$, depending on the applied voltage used. Different W values have been evaluated thanks to simulations performed with TCAD Sentaurus, based on the SRP profiles measured for each laser condition of "standard" samples (notably shown in Fig. 3) [24], and for different applied voltages between -1 and -10 V. $D_m$ were also calculated using phase field simulations, calibrated with experiment data [25].

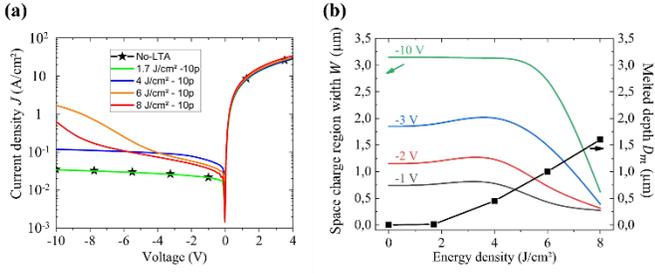

Fig. 4. (a) I-V characteristics for different ED, at 10 pulses. These data were obtained from measurements on 225 µm-radius Schottky diodes at room temperature on "standard" samples. (b) Space charge region width W (TCAD Sentaurus simulations) and melted depth D_m (phase field simulations) as function of the ED for different applied voltage values (between -1 and -10 V).

The obtained results are shown in Fig. 4(b) as a function of the applied ED in comparison to Dm. The space charge region width remains constant up to 4.0 J/cm², independent of the applied voltage. For higher ED (≥ 6.0 J/cm²), a decrease in W is observed, which is caused by the modification of the depth-concentration profile due to the introduction of electrically active oxygen atoms (Fig. 3). W and Dm can be compared at different applied voltages, for example at -1 V and -10 V. For -1 V, the width of the space charge region lies between the melt depths measured at 4.0 and 6.0 J/cm². This means that the melt regions obtained at 1.7 and 4.0 J/cm² are fully probed, whereas those at 6.0 and 8.0 J/cm² are partially probed (~40% and ~15%, respectively). For -10 V, the same behavior is observed and only the diode annealed at 8.0 J/cm² is partially probed (~40%).

Two contributions to the leakage current J can be extracted from our cylindrical Schottky diodes: the perimeter current $J_P$ and the area current $J_A$ [26-28] (2).

$$J = J_A + \frac{P}{A} \times J_P \qquad (2)$$

with the contact area A and perimeter P. $J_P$ is essentially impacted by the defects introduced during the diode fabrication process, while $J_A$ is affected by the defects incorporated during the PLA process. They depend on the area and lateral surface of the cylinder, respectively. Since this study focuses on the impact of the PLA process on the electrical properties of diode structures, only the variations of $J_A$ are considered. Its evolution as a function of the applied ED is plotted in Fig. 5(a), for two reverse bias voltages of -1 and -10 V. Taking into account the comparison between W and $D_m$ made in Fig. 4(b), the data points corresponding to partially and fully probed melted regions using Schottky diodes are colored in red and green, respectively. The corresponding configurations are illustrated in Fig. 5(c). On this basis, different conclusions can be drawn from Fig. 5(a). For both voltages, $J_A$ increases to a maximum value and then decreases. The decreasing part of both curves is consistent with partially probed conditions (red dots), whereas the increasing part fits with the fully probed melt regions (green dots). For the partially probed laser ED conditions, no significant contribution of $J_A$ is measured in the complex impurity-rich region (carbon and oxygen). Therefore, their associated area current density values are largely underestimated and cannot be considered to considered as a link between $J_A$ and ED.

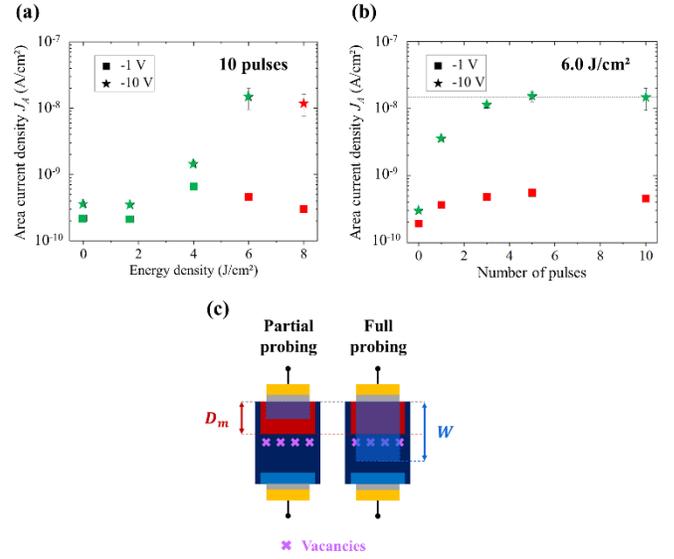

Fig. 5. (a) Area current density $J_A$ as function of the ED for two different voltages (-1 and -10 V), and measured on "standard" samples annealed with 10 pulses using Schottky diodes. (b) Area current density JA as function of the number of pulses for two different voltages (-1 and -10 V), measured on "standard" samples annealed at 6.0 J/cm² using Schottky diodes. Green and red colours refer to fully probed and partially probed measurements, respectively (see Fig. 4(b)). (c) Schematics of Schottky diodes biased at two different voltages and showing partially and fully probed melted regions.

The trend outlined by the green dots shows an increasing area density $J_A$ current with the ED. A maximum area current density value of ~1.3x10$^{-8}$ A/cm² is reached at 6.0 J/cm² (-10 V), which is approximately 50 times higher than the minimum value (~2.5x10$^{-10}$ A/cm²) measured for ED below 2.0 J/cm². Possibly, these $J_A$ variations can be attributed to the increasing incorporation of complex impurities (carbon and oxygen) throughout the entire melt depth. To confirm this hypothesis, the effect of multi-pulse annealing on $J_A$ was evaluated by varying $N_P$ at fixed ED.

In Fig. 5(b) we plot the area current $J_A$ of Schottky diodes annealed at 6.0 J/cm² with 1, 3, 5 and 10 pulses of "standard" samples, measured with an applied voltage of -1 V and -10 V. As indicated by the colors of the data points, voltages of -1 V and -10 V allow a partial and complete probing of the melted layer, respectively. Contrary to Fig. 5(a), both curves exhibit an increasing $J_A$ as successive pulses are added up to a saturation plateau at 5 pulses. The only difference between the two voltages used can be found in the absolute values of $J_A$ which are systematically higher at -10 V than at -1 V. This is particularly the case for the saturation plateaus, which rise to ~5.0x10$^{-10}$ and 1.4x10$^{-8}$ A/cm² for -1 V and -10 V, respectively. This results from the variations of the probed width with the applied voltage (Fig. 5(c)). Firstly, the transient regime observed between 0 and 5 pulses was explained by a slight progressive shift of the l/s interface towards deeper melt depths. Indeed, an increase in the incorporation of impurities can cause a change in the reflectivity index [29,30]. After 5 pulses, when $J_A$ reaches a plateau, it reflects the negligible contribution of point defects (vacancies) at the l/s interface compared to their contribution from the whole melt layer. However, this hypothesis is not consistent with the high measured values of $J_A$ at -10 V and the presence of plateaus.

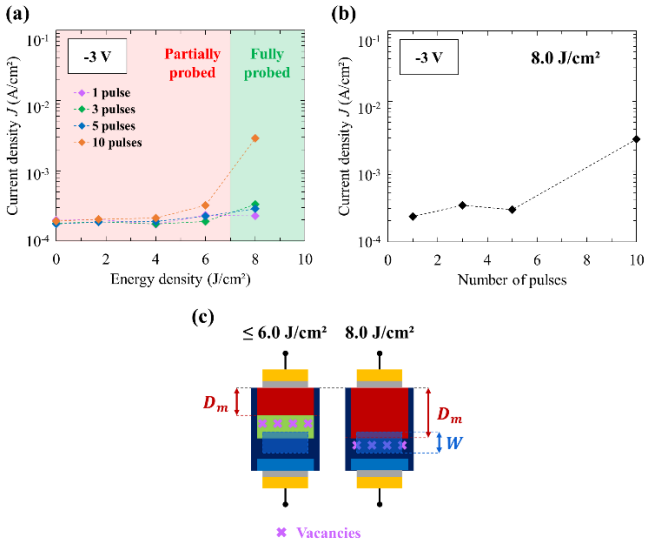

Fig. 6. (a) Current density J as function of the ED for different number of pulses (between 1 and 10 pulses) and measured with PN diodes, under -3 V bias. (b) Current density J as function of the number of pulses at 8.0 J/cm², under -3 V bias. (c) Associated schematics of PN diodes comparing the positioning of the point defects accumulation area (below the melt region) and the probed space charge region W of the structure under -3 V, for ED below than 6.0 J/cm² and at 8.0 J/cm².

If vacancies had a negligible contribution, their impact would not be so clearly highlighted when varying the ED (Fig. 5(a)). Also, the effect of vacancies would start to increase significantly once the l/s interface is reached at 5 pulses, as illustrated in Fig. 1, due to the accumulation effect. Another hypothesis that can be formulated is the existence of a maximum vacancy concentration $V_{max}$, that increases with the temperature. This would explain the increase in $J_A$ when the ED (i.e. the maximum temperature) is increased and its saturation when multiple laser pulses are applied. Indeed, at 6.0 J/cm² and after 5 pulses, not only is $V_{max}$ reached at the l/s interface, but also the vacancy profile in the melted region is completely extended towards the surface. Some vacancy diffusion may occur in the solid phase, but not as much as in the liquid phase. Therefore, increasing the number of pulses would not drastically increase the number of vacancies, hence the observed plateaus.

*2) PN diodes measurements*

PN diodes were fabricated to confirm the vacancies as the main contribution to $J_A$, which increases with the number of pulses. Fig. 6(a) shows the current density J obtained by PN diodes at -3 V as a function of ED, for 1 to 10 pulses. When the ED is equal to or lower than 6.0 J/cm², no clear influence of the ED and the number of pulses on J is observed. In fact, for all these laser conditions, the melt depth is too shallow to reach the PN junction, so the l/s interface (i.e. the vacancy concentration peak) cannot be probed (Fig. 6(c)). At 8.0 J/cm², the melt depth slightly exceeds the PN junction depth so the l/s interface is included in the space charge region W (Fig. 6(c)). Considering this particular configuration, no impact of the number of pulses on J is observed between 1 and 5 pulses, which remains around ~2.5x10⁻⁴ A/cm² (Fig. 6(b)). At 10 pulses, the measured current density is ~3.0x10⁻³ A/cm², which is 10 times higher than between 1 and 5 pulses, confirming the effect of the point defect accumulation on the leakage current. It should be noted that the increase in J does not coincide with the increase in $J_A$, measured with Schottky diodes (Fig. 5(b)). This could be due to several factors. In particular, the presence of boron to fabricate the PN junction could alter the dynamics of vacancy creation and diffusion. Also, the PN configuration could suffer from a lack of signal due to the reduced probed width W compared to the melt depth Dm. Excluding diode structure issues, the abrupt increase in J between 5 and 10 pulses can be explained by the progressive approach of the oxygen impurities close to the l/s interface [11]. Therefore, this measurement suggests that the increasing leakage current densities could originate from the interaction between point defects (vacancies) and complex impurities (at least oxygen). The saturation of $J_A$ shown in Fig. 5(b) indicates that vacancies can be considered as dominant leakage current drivers (i.e. trap centers), although they are in minority.

### C. Impurities interaction

The isolated contributions of complex impurities (oxygen and carbon) and point defects (vacancies) were studied using Schottky diodes. Thanks to PN diodes, we have highlighted an interaction between the two defect types as a plausible source of the leakage current. Indeed, both complex impurities and vacancies, despite their different depth profiles outlined in Fig. 1, can interact within the melted region during their respective diffusion. In our previous study, we showed that the nature of the surface oxide has a strong influence on the in-diffusion species during laser annealing [11]. Indeed, oxygen impurities are observed for both native and thermal oxides whereas carbon impurities are only detected for native oxide. Thus, the quality of the surface oxide can drastically reduce the incorporation of carbon atoms. To evaluate the influence of carbon atoms and their interaction with vacancies on the leakage current, DLTS measurements were performed on both "standard" and "thermal" samples annealed at 6.0 J/cm² with 10 pulses, which are likely to have high concentrations level of impurity complexes and point defects (high ED, multiple number of pulses).

The measured DLTS spectra measured in melted region of these samples are plotted in Fig. 7(a). The spectrum of the "standard" sample displays two broad peaks at 168 K and 252 K, labelled $E_1$ and $E_2$, respectively. For the "thermal" sample, only $E_2$ is observed. The two peaks are attributed to two discrete deep levels in the sensitivity range and their signature is similar to spectrum reported in the literature [30]. Various trap identifications combining vacancies (V), oxygen (O) and carbon (C) impurities have been proposed and are still under discussion: V, V², V5, V-O, V²-O² or V-C-O [31]. These hypotheses all consist of lacunar-type point defects, which may or may not be related to impurities. Based on the conclusions established of our previous study, the $E_2$ signature peak observed in both "thermal" and standard" could be directly and exclusively attributed to the incorporation of oxygen impurities. On the contrary, the $E_1$ signature peak, measured only on the "standard" sample spectrum, must be associated with the in-diffusion of carbon impurities.

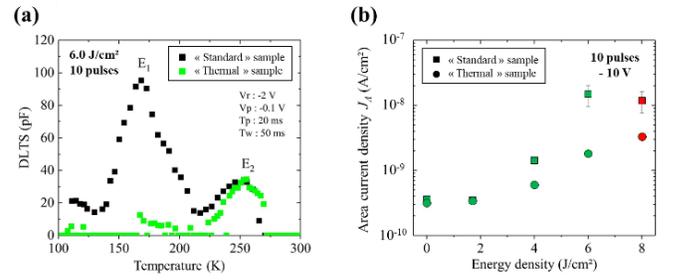

Fig. 7. (a) DLTS-temperature spectra measured with Schottky diodes annealed by laser at 6.0 J/cm² and 10 pulses on both "standard" (native oxide, black squares) and "thermal" (thermal oxide, green squares). (b) Evolution the area component $J_A$ of the leakage current for both "standard" and "thermal" Schottky diodes, measured at -10 V, for multiple laser-annealing conditions (ED between 0 and 8.0 J/cm², 10 pulses). Green and red colors refer to fully probed and partially probed measurements, respectively (see Fig. 4(b)).

Fig. 7(b) shows the area component $J_A$ of the leakage current, extracted from I-V measurements performed with Schottky diodes at -10 V on both "standard" and "thermal" samples, annealed by laser at ED between 0 and 8.0 J/cm² with 10 pulses. For both surface types, and in the fully probed range, we observe a global increase in $J_A$ as the ED is increased. However, distinct trends are observed from 4.0 J/cm², with a more pronounced increase in $J_A$ for the "standard" samples. The growing divergence between the two $J_A$ curves can be attributed to the increasing incorporation of carbon atoms into the melted region with increasing ED, which enhances the impact of trap centers on the leakage current. Thus, the remaining area component of the leakage current $J_A$ in "thermal" samples must be associated with the single combination of vacancies and oxygen impurities, as suggested in the previous section.

To further understand these results, it is interesting to interpret them in terms of energy levels [32-34]. The small active fraction of the incorporated oxygen impurities (< 1%) could have energy levels close to the conduction band, similar to conventional donors such as arsenic or phosphorus. Inactive carbon and oxygen complex impurities, when associated with vacancies, form trap centers (such as V-O, V-C or V-C-O) whose energy levels could be positioned in deep levels, closer to the mid-gap than oxygen impurities, thus facilitating carrier transitions towards between the valence and conduction bands [35].

## IV. Conclusion

In this work, we present a study that evaluates the impact of impurities, such as vacancies and complex impurities, incorporated by high-ED multi-pulse PLA, on the leakage current. The leakage current variations were measured by varying the laser annealing conditions and using dedicated Schottky and PN diodes. To properly analyze the results extracted from the Schottky diodes, the probed region width was calculated using a simulation tool and then, compared with the corresponding melt depths. For each current-voltage measurement, only the contribution related to the introduction of defects during the PLA process, i.e. the area current density $J_A$, was extracted. From these experiments, we conclude that the ED defines the maximum solubility concentration levels of vacancies within the liquid and solid phases of silicon. The multiplication of laser pulses allows the accumulation of vacancies up to the maximum level, defined by the ED. Once reached, the area current density is saturated. The use of PN diodes confirmed the driving role of vacancies on the leakage current, but also revealed a possible enhanced effect when combined with complex impurities such as oxygen. Finally, using DLTS measurements, we confirmed that performing PLA on native oxide-capped silicon favors the incorporation of carbon atoms, in contrast to silicon with thermally stabilized and controlled oxides. In conclusion, high-ED multi-pulse PLA favors the incorporation and the in-diffusion of oxygen and carbon atoms from the surface and external sources, such as oxides or the ambient environment. This process also generates vacancies at the l/s interface that diffuse toward the surface. The interaction between the two defect types leads to the formation to the formation of trap centers, that cause increasing leakage currents, mainly driven by minority vacancies.

## Acknowledgment

This work was financially supported by the Nano2017 French program and was partly supported by LAAS-CNRS micro and nanotechnologies platform, member of the French RENATECH network. We dedicate this article in loving memory of our friend and colleague Fuccio Cristiano, who passed away in January 2024, for his extensive contribution to this work and to other related projects.